\documentclass[12pt]{article}
\usepackage{epsfig}
\begin{document}
% \eqsec  % uncomment this line to get equations numbered by (sec.num)
\title{Elimination of ambiguities in $\pi\pi$ amplitudes using Roy's equations
%ELIMINATION OF AMBIGUITIES IN $\pi\pi$ AMPLITUDES USING ROY'S EQUATIONS
\thanks{Presented at the Meson 2002, Seventh International Workshop on
Production, Properties and Interactions of Mesons, Cracow, Poland, 24-28 May
2002.}
}

\author{R. Kami\'nski$^a$, L. Le\'sniak$^a$ and B. Loiseau$^b$ \\
$^a$ Henryk Niewodnicza\'nski Institute of Nuclear Physics, \\ 
PL 31-342 Krak\'ow, Poland\\
$^b$ LPNHE, Universit\'e P. et M. Curie, 4, Place Jussieu, \\
         75252 Paris Cedex 05, France}

\maketitle

\vspace{-0.8cm}

\begin{abstract}

Roy's equations are used to check if scalar-isoscalar $\pi-\pi$ 
amplitudes fitted to the ``up-flat'' and ``down-flat'' 
phase shift solutions fulfill crossing symmetry below 1 GeV.
It is shown that the amplitude fitted to the ``down-flat'' solution satisfies
crossing symmetry while the ``up-flat'' 
one does not. In such a way the ``up-down'' 
ambiguity in the
scalar-isoscalar phase shifts is resolved in favour of the ``down-flat'' solution.

\end{abstract}

\vspace{-0.1cm}


\vspace{0.15cm}

Direct study of the $\pi\pi$ scattering is beyond present experimental 
possibilities. 
However, phenomenological phase shifts can be obtained through partial wave 
analyses of final states of reactions in which pions are produced.  
These analyses 
are often model dependent and can lead to \mbox{ambiguous results.}

In 1997 a study of the $\pi^-p\uparrow\to\pi^+\pi^-n$ reaction on a 
polarized target was performed for the $m_{\pi\pi}$ effective 
mass between 600 and 1600 MeV leading to four solutions for the 
$\pi\pi$ scalar-isoscalar phase shifts below 1 GeV \cite{kam97}. 
Using the unitarity constraint two ``steep'' solutions were rejected 
while the 
two remaining ones, called ``down-flat'' and ``up-flat'', passed~this~test~\cite{kam00}. 

In order to eliminate this ``up-down'' ambiguity one can 
check if the corresponding amplitudes satisfy crossing symmetry. 
Roy's equations~\cite{roy71} can serve as a tool to perform this check and to
correlate 
scalar-isoscalar, scalar-isotensor and vector-isovector phase shifts
determined near the $\pi\pi$ threshold and at higher energy.
Experimentally determined phase shifts $\delta$ and 
inelasticities $\eta$ are used to calculate the
imaginary parts of the amplitudes which can be inserted  
into Roy's equations.
The resulting real parts of the amplitudes, called ``out'' can be
compared with the real parts, $\eta(\sin2\delta)/2$, 
 directly calculated from the phase shifts 
and called ``in''.
We check the quantitative agreement between the ``in'' and ``out'' real parts
to test how well a given set of amplitudes satisfies Roy's equations~for~$m_{\pi\pi}<970$~MeV.

A comprehensive analysis of Roy's equations with  
a special emphasis  on the $m_{\pi\pi}$ range from  
 threshold to 0.8 GeV has recently appeared~\cite{anan}. 
Here we pay particular
attention to the range between 0.8 and 1 GeV where the largest differences 
between the ``up-flat'' and ``down-flat'' solutions occur. 
%reaching values up to 45$^{o}$ (see Fig. 4 in \cite{kam00}). 
Below 970 MeV we parameterize the amplitudes corresponding to the 
``down-flat'' and ``up-flat'' data using Pad\'e's 
approximants with 8 
free parameters.
Above 970 MeV we use the two corresponding amplitudes A and C of
our analysis with three coupled channel interactions
($\pi\pi,\ K\bar K$ and an effective $4\pi$ system) 
\cite{kam99}.
%This three channel model yields particularly good fit to the ``down-flat'' data 
%from 600 MeV to 1600 MeV (fit A) but the ``up-flat'' data were reasonably well 
%described only above 970 MeV (fit C). 
At 970 MeV the Pad\'e  and the A or C amplitudes are smoothly
connected by a proper choice of two parameters in Pad\'e's 
formula.
%Higher energy contributions ($s\geq\Lambda$) of all partial waves and 
%lower energy contributions ($s\leq\Lambda$) are included in so called
%driving terms.
The remaining 6 parameters are obtained in fitting both experimental data 
and Roy's equations up to 970 MeV.
Near the $\pi\pi$ threshold we use the data of 
\cite{E865} and above 600 MeV the
 ``up-flat'' and
 ``down-flat'' phase shifts of \cite{kam97}. 
The $S$-wave isotensor amplitude is parameterized using the rank-two
potential model
\cite{kam99}
with 4 parameters fitted to 
the data set A of \cite{hoogland74}.
The $P$-wave amplitude is parameterized as in \cite{anan} and its 5 parameters
are fitted to data and to satisfy Roy's equations.

For the ``down-flat'' case we have obtained the isoscalar amplitude describing
well the
experimental data ($\chi^2 = 16$ for 18 points) and simultaneously fulfilling
Roy's equations.
The differences between the real part values ``in'' and ``out''
were smaller than $10^{-4}$ for all three $\pi\pi$ amplitudes.
In the ``up-flat'' case such a fit could not be obtained since the above differences for
the isoscalar wave were as large as 0.2 around 900 MeV.

We have studied the influence of the experimental errors on the $\pi\pi$ input 
amplitudes by calculating Roy's equations for two extreme isoscalar amplitudes fitted to the 
data points shifted upwards (upper ``in'') or downwards (lower ``in'') by their errors.
Below 600 MeV these fits are constrained to approximate the previously obtained ``down-flat'' 
amplitude and in particular to reproduce its scattering length $a_{0}^{0} = 0.224$ and
the slope parameter $b_{0}^{0} = 0.272$.
As seen in Fig. 1, in the ``down-flat'' case, both ``out'' curves lie 
inside the band limited by ``in'' curves up to 930 MeV.
%In the ``up-flat'' case above 850 MeV, the upper
% ``out'' curves lie largely outside the ``in'' band.
On the contrary, in the ``up-flat'' case above 850 MeV, the ``out'' band
 lies outside~the~``in''~band.

We conclude that the amplitude corresponding to the ``down-flat'' solution 
does fulfill Roy's equations but the ``up-flat'' amplitude does not.
Thus one should accept the ``down-flat'' solution as the physical one and
reject the  ``up-flat'' solution.
In this way the ``up-down'' ambiguity is resolved in favour of the ``down-flat''
solution as it has also been shown in a recent joint analysis of the $\pi^+\pi^-$ and 
$\pi^0\pi^0$ data
\cite{kam01}.

\vspace{-0.25cm}

\begin{figure}[htbp]\centering   
\mbox{\epsfxsize 11.5cm\epsfysize 5.1cm\epsfbox{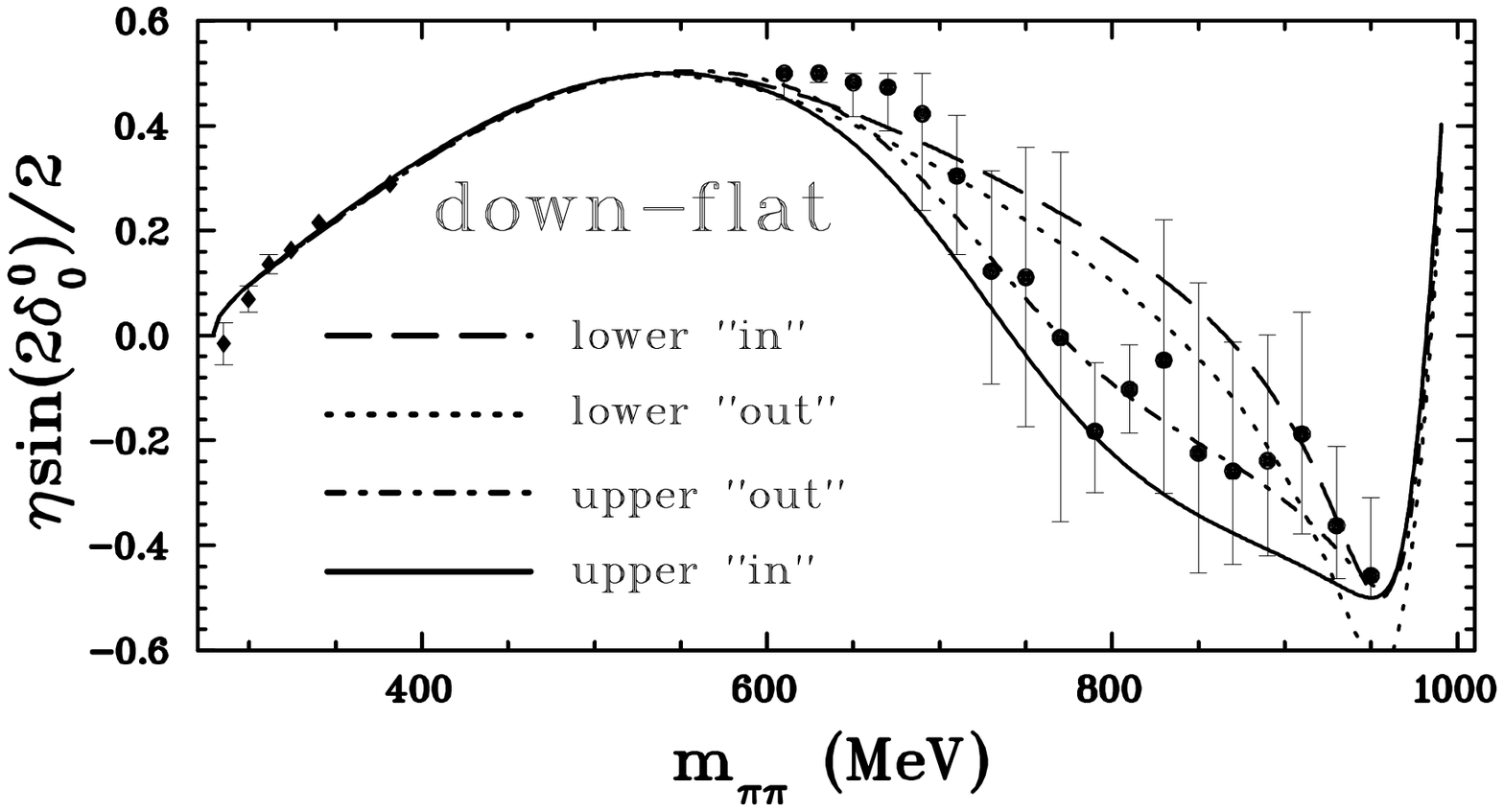}}\\   
\mbox{\epsfxsize 11.5cm\epsfysize 5.1cm\epsfbox{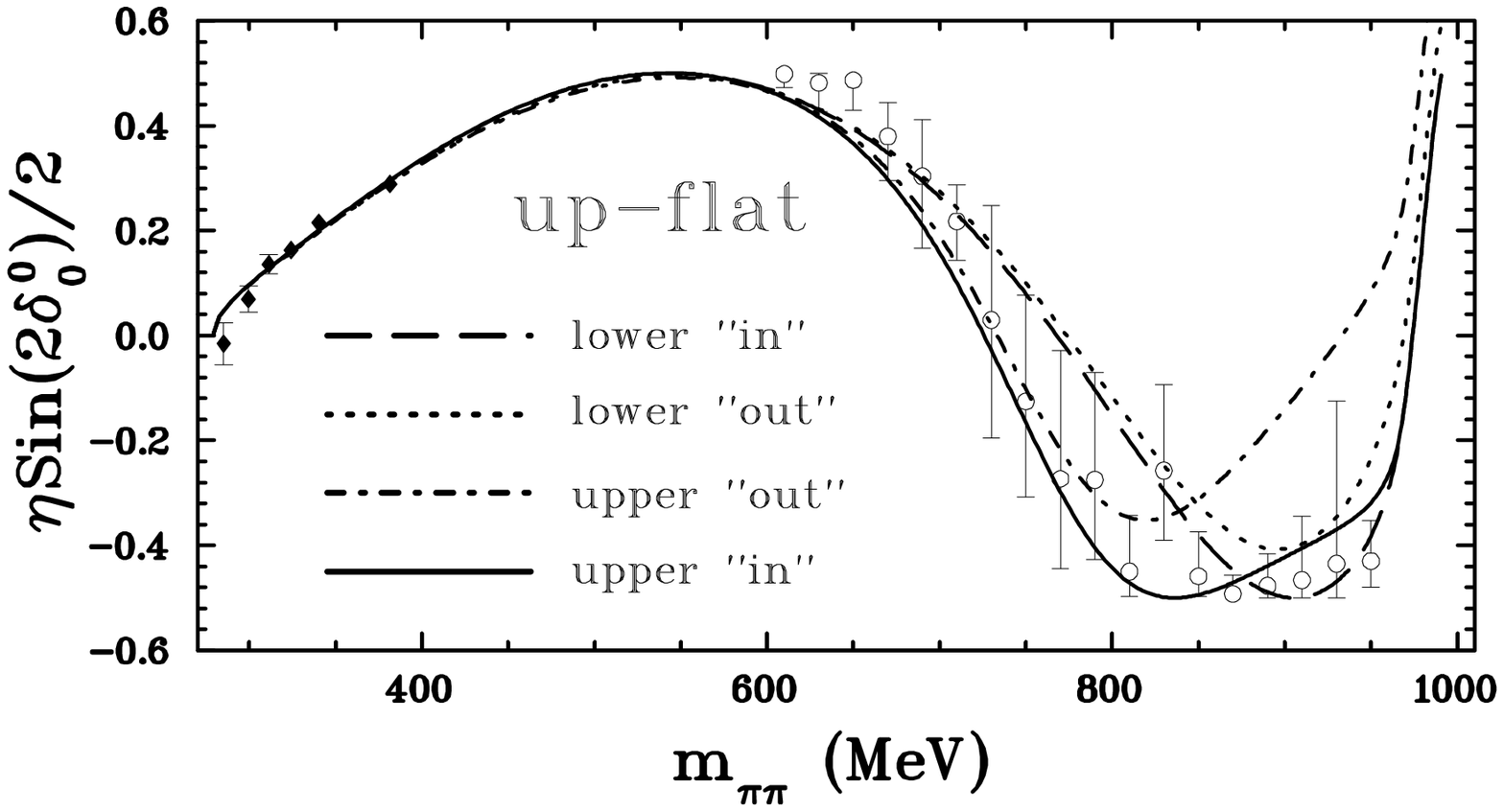}}\\   
{Fig. 1: Input bands (solid and dashed lines) and output bands 
(dotted and dot-dashed lines) computed from Roy's equations for the $S$-wave
isoscalar amplitude.
Diamonds denote the data of \cite{E865} and circles those of \cite{kam97}.   
}    
\end{figure} 

\vspace{-0.1cm}

{\it Acknowledgments}: This work was supported by IN2P3-Polish laboratories
 Convention (project No. 99-97). LPNHE is Unit\'e de Recherche des Universit\'es
Paris 6 et Paris 7, associ\'ee au CNRS.


\vspace{-0.6cm}
  
\begin{thebibliography}{99}

\vspace{-0.2cm}

    \bibitem{kam97}  R. Kami\'nski, L. Le\'sniak, K. Rybicki, Z. Phys. 
     \textbf{C74} (1997) 79.

    \bibitem{kam00}  R. Kami\'nski, L. Le\'sniak, K. Rybicki, Acta Phys. 
    Pol. \textbf{B31} (2000) 895.

    \bibitem{roy71}  S.M. Roy, Phys. Lett. \textbf{36B} (1971) 353,
     Helv. Phys. Acta \textbf{63} (1990) 627.

    \bibitem{anan}  B. Ananthanarayan {\it et al.},
%G. Colangelo, J. Gasser, H. Leutwyler, 
Phys. Rep. \textbf{353} (2001) 207.

    \bibitem{kam99} R. Kami\'nski, L. Le\'sniak, B. Loiseau, 
Phys. Lett. \textbf{B413} (1997) 130. 
%Eur. Phys. J. \textbf{C9} (1999) 141.

\bibitem{E865} S. Pislak {\it et al.}, Phys. Rev. Lett. {\bf 87} (2001) 221801.

    \bibitem{hoogland74} W. Hoogland {\it et al.},  
%\textbf{B69} (1974) 266 ; 
Nucl. Phys. \textbf{B126} (1977) 109.

    \bibitem{kam01} R. Kami\'nski, L. Le\'sniak, K. Rybicki, {\bf hep-ph}/0109268, 
accepted for publication in Eur. Phys. J. C (2002).

\end{thebibliography}
\end{document}